\documentclass[prb,showpacs,twocolumn,amsmath,amssymb]{revtex4}
\usepackage{graphicx}
\usepackage{bm}

\def\br{\bm{r}}
\def\bk{\bm{k}}
\def\bq{\bm{q}}
\def\bQ{\bm{Q}}
\def\bp{\bm{p}}
\def\bv{\bm{v}_{\bk}}
\def\im{\,\mathrm{Im}\,}
\def\re{\,\mathrm{Re}\,}

\begin{document}

\title{Quantum fluctuations in Larkin-Ovchinnikov-Fulde-Ferrell superconductors}

\author{K. V. Samokhin$^{1}$ and M. S. Mar'enko$^{1,2}$ }

\affiliation{$^{1}$ Department of Physics, Brock University,
St.Catharines, Ontario, Canada L2S 3A1\\
$^{2}$Department of Physics and Astronomy, Hofstra University,
Hempstead, New York 11549, USA}
\date{\today}

\begin{abstract}
We study the superconducting order parameter fluctuations near the
phase transition into the Larkin-Ovchinnikov-Fulde-Ferrell state
in the clean limit at zero temperature. In contrast to the usual
normal metal-to-uniform superconductor phase transition, the
fluctuation corrections are dominated by the modes with the wave
vectors away from the origin. We find that the superconducting
fluctuations lead to a divergent spin susceptibility and a
breakdown of the Fermi-liquid behavior at the quantum critical
point.
\end{abstract}


\maketitle

\section{Introduction}
\label{sec: intro}

Magnetic field suppresses superconductivity, regardless of the
pairing symmetry, via the coupling to the orbital motion of
electrons in the Cooper pairs.\cite{Tink96} In spin-singlet
superconductors, the pairs are also broken by the Zeeman
interaction of electron spins with an applied field $\bm{H}$, or
by the exchange interaction with localized spins in a magnetic
crystal, which is known as the paramagnetic, or Pauli, mechanism.
If the orbital effects are neglected, then as shown by Larkin and
Ovchinnikov\cite{LO64} and Fulde and Ferrell\cite{FF64} (LOFF),
the competition between the paramagnetic pair breaking and the
condensation energy results in the formation at low temperatures
of a peculiar non-uniform superconducting state with a periodic
modulation of the order parameter, whose critical field $H_c$
exceeds the Clogston-Chandrasekhar paramagnetic limit for a
uniform state.\cite{Clog62,Chan62} The superconducting order
parameter in the simplest realizations of the LOFF state is either
a single plane wave: $\eta(\br)=\eta_0e^{i\bq_c\br}$, or a
superposition of two plane waves: $\eta(\br)=\eta_0\cos\bq_c\br$,
where $\bq_c$ is the wave vector of the LOFF instability. In
general, the order parameter structure can be more complicated and
is determined by minimizing the nonlinear Ginzburg-Landau free
energy.

For a long time the LOFF state had been considered a theoretical
curiosity, because its experimental detection required the
fulfillment of some rather stringent conditions. First, the
orbital effects are detrimental to the LOFF state and therefore
should be weak enough. The relative importance of the orbital and
spin pair-breaking mechanisms is measured by the Maki parameter
$\alpha_M=\sqrt{2}H^{(0)}_{c2}/H_{CC}$,\cite{Maki64} where
$H^{(0)}_{c2}$ is the upper critical field in the absence of spin
interactions, and $H_{CC}$ is the Clogston-Chandrasekhar critical
field. In the pure paramagnetic limit, the orbital pair-breaking
is absent altogether and $\alpha_M=\infty$. The orbital effects
were included in the LOFF model with a spherical Fermi surface by
Gruenberg and Gunther,\cite{GG66} who found that at $T=0$ and
$\alpha_M\gtrsim 1.8$ the order parameter below the upper critical
field is modulated along the direction of the applied field. The
coordinate dependence of the pair wave function in the transverse
directions is described by the lowest Landau level, i.e. is the
same as in the pure orbital case.\cite{Abr57,HW64} There is
another phase transition at a lower field, either into the usual
mixed state or into the uniform superconducting state at
$\alpha_M=\infty$, resulting in the appearance of a characteristic
wedge-like region in the $H-T$ phase diagram at low temperatures
and high fields. In most ``classical'' superconductors, however,
$\alpha_M\lesssim 1$, so the orbital pair-breaking dominates and
the LOFF state is never realized.

One possible way to reduce the orbital effects was proposed by
Bulaevskii,\cite{Bul74} who pointed out that in a layered
superconductor with the electron orbital motion confined to the
layers, the Maki parameter depends on the angle $\theta$ between
the direction of $\bm{H}$ and the layers, making the paramagnetic
effects dominant in a narrow angle interval near $\theta=0$. As
$\theta$ approaches zero, the system undergoes a series of phase
transitions between the LOFF states corresponding to successive
higher Landau levels. The full $H-T$ phase diagram was worked out
in Ref. \onlinecite{SR97}. At the parallel field orientation there
is no orbital effects, all pair breaking is entirely paramagnetic,
and the region of existence of the non-uniform state turns out to
be larger than in the isotropic 3D case, see also Ref.
\onlinecite{ADF74}. The same ideas also apply to thin
superconducting films,\cite{Fulde73} or to surface
superconductors\cite{BG02}, in parallel fields.

Another obstacle to the experimental realization of the
non-uniform state is its sensitivity to the presence of disorder.
It was found by Aslamazov\cite{Aslam68} in the isotropic case that
the LOFF critical field decreases rapidly with increasing
non-magnetic impurity scattering and eventually becomes smaller
than $H_{CC}$, resulting in the restoration of a first-order phase
transition into the uniform superconducting state. Later the
analysis was extended to the layered case in Ref.
\onlinecite{BG76}, with essentially the same conclusions.

Thus the LOFF state can potentially be observed only if the
superconductor is both paramagnetically limited and sufficiently
clean. These requirements can be simultaneously met in
heavy-fermion compounds, in which $H^{(0)}_{c2}$ is inherently
high due to a short coherence length. Earlier candidates for
hosting the LOFF state included
UPd$_2$Al$_3$,\cite{Gloos93,Modler96} UBe$_{13}$,\cite{Thomas95}
and CeRu$_2$.\cite{Modler96} The odds of finding the LOFF state
are even greater in quasi-low-dimensional superconductors, in
which the orbital pair breaking is
reduced.\cite{BR94,Dupuis95,Shima97,HBBM02} Several experiments on
organic\cite{Single00,MK00,Tanatar02} and cuprate\cite{OBrien00}
superconductors have revealed the features in the the $H-T$ phase
diagram, such as an upturn of the upper critical field and the
presence of an additional phase transition below $H_{c2}$ at low
temperatures, that could be interpreted as signatures of the LOFF
state. Similar features have been recently reported in another
heavy-fermion compound, CeCoIn$_5$.\cite{Rado03,Bianchi03} In all
the cases mentioned above, the spin splitting of the electron
energies in the Cooper pairs was due to the Zeeman interaction
with an applied magnetic field. It was argued in Ref.
\onlinecite{Pickett99} that the LOFF state can be created by the
intrinsic exchange band splitting in the ferromagnetic
superconductor RuSr$_2$GdCu$_2$O$_8$.

Another intriguing possibility of the experimental realization of
the LOFF state has been discussed very recently in the context of
ultracold atomic Fermi gases, such as ${}^{40}$K and ${}^6$Li. By
making the populations of atoms in two different hyperfine states
unequal, one controls the mismatch between their Fermi
surfaces.\cite{Zwier05,Part05} When the pairing interaction
between the two fermion species is turned on, the system becomes
formally equivalent to a neutral superconductor in a Zeeman field.
Due to the absence of both the orbital effects and impurities,
this seems to be the most promising setup to study the
paramagnetic pair breaking, including the non-uniform
states.\cite{Comb01,MMI05,SR05,SMPM05,DOR05}

Despite the lack of unambiguous experimental evidence, the LOFF
state has remained a subject of intensive theoretical
investigations in the past decades. In addition to the studies
cited above, we would like to mention Refs.
\onlinecite{BK97,AY01}, in which the Ginzburg-Landau theory was
developed in the pure paramagnetic case in the vicinity of the
tricritical point in the $H-T$ phase diagram, see Fig. 2 below,
where the sign change of the second-order gradient term in the
free energy signals the onset of the non-uniform instability. The
vortex structure in the mixed LOFF state was studied in Refs.
\onlinecite{KRS00,HB01,AI03,YMcD04}. The LOFF model has also been
extended to unconventional, in particular $d$-wave, pairing
symmetries.\cite{Shima97,Sam97,YS98,VSG05} An analysis of the
spatial structure of the LOFF state immediately below the upper
critical field in the isotropic 3D case was done by Larkin and
Ovchinnikov,\cite{LO64} who showed that it is the ``striped''
phase with $\eta(\br)=\eta_0\cos\bq_c\br$ that is energetically
favored at $T=0$. The zero-temperature phase transition from the
normal state was found to be second order, but becomes first order
as temperature increases.\cite{MR71,MHNN98,CT04} On the other
hand, it was argued in Refs. \onlinecite{BR02,MC04} that the phase
transition is always first order below the tricritical point, and
that the order parameter at $T=0$ is represented by a sum of three
cosines.

Another open question concerns the nature of the lower phase
transition separating the LOFF state from the conventional uniform
superconducting state. The only cases studied so far assumed a
one-dimensional periodicity of the order parameter in a purely
paramagnetic and isotropic 2D\cite{BR94,VSG05} or 3D\cite{MHNN98}
system. As the field decreases, the non-linear effects add higher
harmonics to the LOFF state, which starts to resemble a periodic
array of Bloch domain walls separating the regions where the order
parameter is almost uniform. When one approaches the lower
critical field, the period of the domain structure diverges,
indicating a second order phase transition.

While an extensive literature exists about the mean-field
properties of the LOFF state, the superconducting fluctuation
effects have received comparatively little attention. The phase
fluctuations of the non-uniform order parameter were considered by
Shimahara,\cite{Shima98} who found that they are able to destroy
even the quasi-long-range order for the striped LOFF states in the
isotropic 2D case at $T>0$. He also conjectured that in the
isotropic 3D case the long-range order is replaced at finite
temperatures by a quasi-long-range order. This is consistent with
the findings of Ref. \onlinecite{Ohashi02}, where it was shown
that the thermal fluctuations suppress the second order phase
transition into the LOFF state in spatially isotropic systems.
These effects are analogous to the fluctuation-driven destruction
of crystalline order with one-dimensional density
modulation.\cite{LL-vol5} In general, one can expect that, since
the wave vectors of important fluctuating modes in the isotropic
LOFF state are close to a sphere (or a circle in 2D) of radius
$|\bq_c|\neq 0$, the fluctuation effects on observable quantities
will be considerably enhanced compared to the uniform case due to
the increased phase volume of the fluctuations.\cite{Braz75} The
fluctuation effects might still be significant even when the
degeneracy manifold of the LOFF states is reduced to a set of
isolated points in the momentum space: It was recently argued in
Ref. \onlinecite{DY04} that the thermal fluctuations in quasi-2D
$d$-wave superconductors are strong enough for the LOFF phase
transition to become of first order.

In all the works mentioned above only finite temperatures were
considered, in which case the order parameter fluctuations are
predominantly classical. Formally, the classical limit corresponds
to setting the frequency in the fluctuation propagator to zero,
see Sec. \ref{sec: general} below. The focus of the present work
is on the fluctuation effects at $T=0$ above the quantum phase
transition from the normal state to the LOFF state driven by an
external magnetic field. In this case, the dynamic nature of the
superconducting fluctuations cannot be neglected. We assume that
the system can be described by the Bardeen-Cooper-Schrieffer (BCS)
model and also that the quantum LOFF transition is of second
order. We do not include impurities and the orbital effects,
expecting our results to be applicable either to paramagnetically
limited superconductors, or to the Fermi gases of ultracold atoms.
We would like to note that while thermal fluctuations in
superconductors have been actively studied for a long time, see
Ref. \onlinecite{LV-book}, the quantum fluctuations at low
temperatures have only recently become a subject of theoretical
investigation.\cite{RC97,MS01,GL01,GDS03,LSV05}

The paper is organized as follows: In Sec. \ref{sec: general}, we
derive the general expression for the fluctuation propagator in
the normal state above the LOFF phase transition. We consider both
the isotropic case, in which the LOFF states are infinitely
degenerate in the momentum space, and the generic case, in which
the degeneracy is lifted due to the band structure and/or the gap
anisotropy. In Sec. \ref{sec: corrections}, we calculate the
quantum fluctuation corrections to the spin susceptibility and to
the decay rate of fermionic quasiparticles.

\section{LOFF Fluctuation propagator}
\label{sec: general}

We consider a clean spin-singlet BCS superconductor in an external
magnetic field $\bm{H}$. The coupling of the electron charges to
the vector potential is neglected, so that the superconductivity
is affected by the field only through the Zeeman splitting of the
single-particle bands. The Hamiltonian is given by
\begin{eqnarray}
\label{H}
    &\displaystyle H=\sum_{\bk}(\xi_{\bk}\delta_{\alpha\beta}-
     h\sigma_{3,\alpha\beta})c^\dagger_{\bk\alpha}c_{\bk\beta}&\nonumber\\
    &\displaystyle  +\sum_{\bq,\bk_{1,2}}V_{\bk_1\bk_2}(\bq)
     c^\dagger_{\bk_1+\frac{\bq}{2},\uparrow}c^\dagger_{-\bk_1+\frac{\bq}{2},\downarrow}
     c_{-\bk_2+\frac{\bq}{2},\downarrow}c_{\bk_2+\frac{\bq}{2},\uparrow}.&
\end{eqnarray}
The first term here is the free-fermion part, where
$\xi_{\bk}=\epsilon_{\bk}-\mu$, $\epsilon_{\bk}$ is the band
dispersion, $\mu$ is the chemical potential,
$\alpha,\beta=\uparrow,\downarrow$ is the spin projection on the
quantization axis along $\bm{H}$, $h=\mu_B H$ is the Zeeman field,
$\mu_B$ is the Bohr magneton, and $\sigma_3$ is the Pauli matrix
(we use the units in which $\hbar=k_B=1$, and assume that the
Land\'e factor $g=2$). The Hamiltonian (\ref{H}) can also be
applied to a ferromagnetic superconductor in zero applied field,
in which case the electron bands are split due to the exchange
interaction with the magnetically ordered localized spins.

The second term in Eq. (\ref{H}) is the pairing interaction, which
is effective only in the vicinity of the Fermi surface defined by
the equation $\xi_{\bk}=0$, i.e. at
$|\xi_{\pm\bk_{1,2}+\bq/2}|\leq\omega_{\max}$, where
$\omega_{\max}$ is the BCS energy cutoff. We make a simplifying
assumption that the interaction matrix can be factorized:
\begin{equation}
\label{interaction}
    V_{\bk_1\bk_2}(\bq)=-\lambda(\bq)\phi_{\bk_1}\phi_{\bk_2},
\end{equation}
where $\phi_{\bk}=\phi_{-\bk}$ is the symmetry factor, which is
nonzero only inside the BCS shell, i.e. at
$|\xi_{\bk}|\leq\omega_{\max}$. The symmetry factor is assumed to
be real and normalized: $\langle\phi_{\bk}^2\rangle=1$, where the
angular brackets stand for the Fermi-surface average. In the
group-theoretical language, $\phi_{\bk}$ is the basis function of
an even one-dimensional irreducible representation $\Gamma$ of the
point group ${\cal G}$ of the crystal, which can have zeros,
symmetry-imposed or accidental, somewhere on the Fermi surface.
The pairing is said to be conventional if $\Gamma$ is the unity
representation, and unconventional otherwise.\cite{Book} To make
sure that the energies of all four fermions participating in the
BCS interaction are less than $\omega_{\max}$, one has to further
assume that the function $\lambda(\bq)$ is nonzero only if
$|\bq|\leq q_{\max}\sim\omega_{\max}/v_F$, where $v_F$ is the
Fermi velocity. In the calculations below we replace
$\lambda(\bq)$ by a coupling constant $\lambda>0$, introducing an
explicit momentum cutoff in $\bq$-integrals if needed.

The order parameter dynamics in the normal state is described by
the fluctuation propagator\cite{LV-book}
\begin{equation}
\label{L average}
     {\cal L}(\bq,\nu_m)=\frac{1}{\lambda^{-1}-{\cal C}(\bq,\nu_m)},
\end{equation}
where $\nu_m=2\pi mT$ is the bosonic Matsubara frequency, and
${\cal C}(\bq,\nu_m)$ is the particle-particle propagator (the
Cooperon), see Fig. 1. Calculating the diagrams, we obtain:
\begin{eqnarray}
\label{L}
    &&\frac{1}{N_F}{\cal L}^{-1}(\bq,\nu_m)=\ln\frac{T}{T_{c0}}
    -\Psi\left(\frac{1}{2}\right)\nonumber\\
    &&\qquad+\left\langle\phi_{\bk}^2\re\Psi\left(\frac{1}{2}+
    \frac{iW_{\bk}+|\nu_m|}{4\pi T}\right)\right\rangle,
\end{eqnarray}
where $N_F$ is the density of states per one spin projection at
the Fermi level, $\Psi(x)$ is the digamma function,
$\Psi(1/2)=-\ln(4e^\mathbb{C})$, $\mathbb{C}\simeq 0.577$ is
Euler's constant,
$T_{c0}=(2e^\mathbb{C}/\pi)\omega_{\max}e^{-1/N_F\lambda}$ is the
zero-field critical temperature of the uniform superconducting
state,
\begin{equation}
\label{W def}
    W_{\bk}=\xi_{\bk+\frac{\bq}{2}}-\xi_{\bk-\frac{\bq}{2}}-2h=
    \bv\bq-2h+O(\bq^3),
\end{equation}
and $\bv=\nabla_{\bk}\xi_{\bk}$ is the quasiparticle velocity at
the Fermi surface.  The cutoff $\omega_{\max}$ has been eliminated
by adding and subtracting the Cooperon at $h=\nu_m=\bq=0$.

\begin{figure}
    \label{fig: Propagators}
    \includegraphics[width=7.5cm]{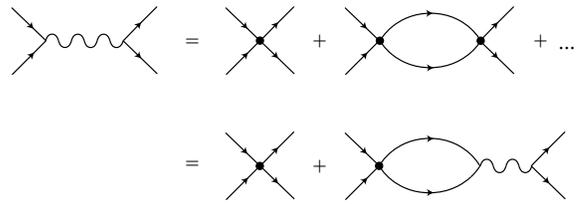}
    \caption{The diagrammatic representation of the fluctuation
    propagator. The solid lines are the Green's functions of fermions
    in the normal state, the bold dots denote the BCS pairing potential.}
\end{figure}

The expression (\ref{L}) is valid at all temperatures, for any
pairing symmetry and band structure [we will keep only the linear
in $\bq$ term in the expansion (\ref{W def}), assuming that the
band structure is such that the higher-order terms are
negligible]. The solution $T(\bq,h)$ of the equation ${\cal
L}^{-1}(\bq,0)=0$ determines the temperature at which the
superconducting instability with the wave vector $\bq$ develops in
a given field $h$. Setting $\nu_m=\bq=T=0$, one finds that the
second-order quantum phase transition into a uniform
superconducting state occurs at
$h_0=(\pi/2e^\mathbb{C})T_{c0}=\Delta_0/2\simeq 0.88T_{c0}$, where
$\Delta_0$ is the BCS gap at $T=0$. In general, the critical
temperature vs field $T_c(h)$, or inversely the critical field vs
temperature $h_c(T)$, can be found by maximizing $T(\bq,h)$ with
respect to $\bq$. According to Refs. \onlinecite{LO64,FF64}, in a
clean isotropic superconductor at $T<T^*\simeq 0.56T_{c0}$ the
maximum of the critical field is achieved at $\bq_c\neq 0$. The
generic phase diagram of a LOFF superconductor is sketched in Fig.
2.

In the vicinity of the critical field $h_c(T)$, the most divergent
contributions to physical quantities come from the low-frequency
fluctuations, so the inverse fluctuation propagator (\ref{L}) can
be expanded in powers of $\nu_m$:
\begin{equation}
\label{L expansion}
     \frac{1}{N_F}{\cal L}^{-1}(\bq,\nu_m)=A(\bq,h)+\tilde A(\bq,h)\frac{|\nu_m|}{2h}
     +O(\nu_m^2),
\end{equation}
where
\begin{eqnarray}
\label{A0 gen}
    A(\bq,h)&=&\left\langle\phi_{\bk}^2\re\Psi\left(\frac{1}{2}+
    \frac{iW_{\bk}}{4\pi T}\right)\right\rangle\nonumber\\
    &&-\Psi\left(\frac{1}{2}\right)+\ln\frac{T}{T_{c0}},\\
\label{tilde A0 gen}
    \tilde A(\bq,h)&=&\frac{h}{2\pi T}\left\langle\phi_{\bk}^2\re\Psi^\prime
    \left(\frac{1}{2}+\frac{iW_{\bk}}{4\pi
    T}\right)\right\rangle.
\end{eqnarray}

\begin{figure}
    \label{fig: Phase Diagram}
    \includegraphics[width=5.6cm]{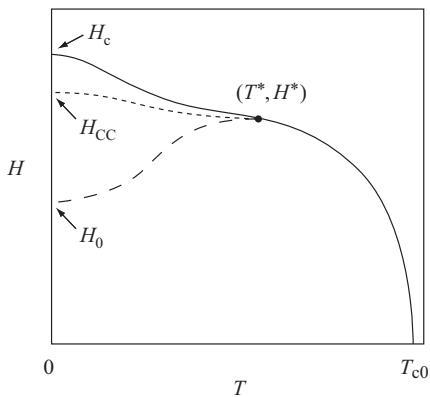}
    \caption{The generic phase diagram of the LOFF superconductor.
    The LOFF state appears at $H=H_c$ (solid
    line) at temperatures
    below the tricritical point $(T^*,H^*)$.
    The dashed lines correspond to the normal metal-to-uniform superconductor phase
    transitions: the first-order Clogston-Chandrasekhar
    transition at $H_{CC}$ (dashed line), and the second-order transition at $H_0$ (long-dashed line).
    The critical field
    of the LOFF state-to-uniform superconductor transition is not shown.}
\end{figure}

We focus on the fluctuation effects at $T=0$, when the explicit
temperature dependence of the expressions (\ref{A0 gen}) and
(\ref{tilde A0 gen}) can be eliminated by using the asymptotic
form of the digamma function: $\Psi(x)=\ln x+O(x^{-1})$ at
$x\to\infty$ (Ref. \onlinecite{AS65}):
\begin{eqnarray}
\label{A clean}
    A(\bq,h)&=&\ln\frac{h}{h_0}+F(\bQ),\\
\label{tilde A clean}
    \tilde A(\bq,h)&=&\tilde F(\bQ),
\end{eqnarray}
where $\bQ=\bq/2h$, and
\begin{eqnarray}
\label{FQ}
    F(\bQ)&=&\left\langle\phi_{\bk}^2\ln|\bv\bQ-1|\right\rangle,\\
\label{tilde FQ}
    \tilde F(\bQ)&=&\pi\left\langle\phi_{\bk}^2
    \delta(\bv\bQ-1)\right\rangle
\end{eqnarray}
If the function $F(\bQ)$ has a minimum at $\bQ=\bQ_c$, then the
upper critical field is given by $h_c=h_0e^{-F(\bQ_c)}$, and the
equilibrium wave vector of the LOFF structure is
$\bq_c=2h_c\bQ_c$.

In the standard theory of superconducting fluctuations, see Ref.
\onlinecite{LV-book}, it is assumed that the maximum of the
critical temperature, or of the critical field, is achieved for
the uniform superconducting state, which makes it possible to
expand ${\cal L}^{-1}(\bq,\nu_m)$ in the vicinity of the origin in
the momentum space. In contrast, the most important critical
fluctuations in the LOFF state have the wave vectors near
$\bq_c\neq 0$. Assuming that $F(\bQ)$ can be expanded in the
Taylor series near the minimum, we obtain the following expression
for the fluctuation propagator:
\begin{eqnarray}
\label{L general}
    &&{\cal L}(\bq,\nu_m)\nonumber\\
    &&\quad=\frac{1}{N_F}\frac{1}{\epsilon+\gamma|\nu_m|+
    K_{ij}(q_i-q_{c,i})(q_j-q_{c,j})},\quad
\end{eqnarray}
where
\begin{equation}
\label{epsilon def}
    \epsilon=\frac{h-h_c}{h_c}
\end{equation}
measures the distance to the quantum critical point,
\begin{equation}
\label{gamma def}
    \gamma=\frac{\tilde F(\bQ_c)}{2h_c},
\end{equation}
and $K_{ij}=(1/8h_c^2)\nabla_i\nabla_jF(\bQ_c)$, $i,j=x,y,z$. In
should be noted that in some cases the frequency expansion (\ref{L
expansion}) of the inverse fluctuation propagator does not exist,
see Sec. \ref{sec: isotropic 2D} below.

\subsection{Isotropic 3D case}
\label{sec: isotropic 3D}

Explicit expressions for $\gamma$ and $K_{ij}$ can only be
obtained in few cases, including a 3D parabolic band,
$\xi_{\bk}=\bk^2/2m-\mu$, with isotropic pairing. In this case a
straightforward integration in Eqs. (\ref{FQ},\ref{tilde FQ})
gives
\begin{eqnarray}
\label{FQ 3D}
    F(\bQ)&=&\frac{1}{2}\ln|x^2-1|+\frac{1}{2x}\ln\left|\frac{x+1}{x-1}\right|-1,\\
    \tilde F(\bQ)&=&\frac{\pi}{2x}\theta(x-1)
\end{eqnarray}
where $x=v_FQ=v_Fq/2h$. The function (\ref{FQ 3D}) has a minimum
at $x=x_c\simeq 1.20$, with $F(\bQ_c)\simeq -0.41$, see Fig. 3.
Thus the quantum phase transition occurs at $h_c\simeq
1.51h_0\simeq 0.75\Delta_0$, and at $h<h_c$ the superconducting
order parameter is spatially modulated, with the wave vector
$q_c=2x_ch_c/v_F\simeq 0.51\xi_0^{-1}$, where $\xi_0=v_F/2\pi
T_{c0}$ is the BCS coherence length. The LOFF critical field $h_c$
exceeds not only $h_0$, but also the Clogston-Chandrasekhar field
$h_{CC}=\Delta_0/\sqrt{2}$, which corresponds to a first-order
phase transition between the normal and the uniform
superconducting states,\cite{Clog62,Chan62} see Fig. 2.

\begin{figure}
    \label{fig: F-3D}
    \includegraphics[width=8cm]{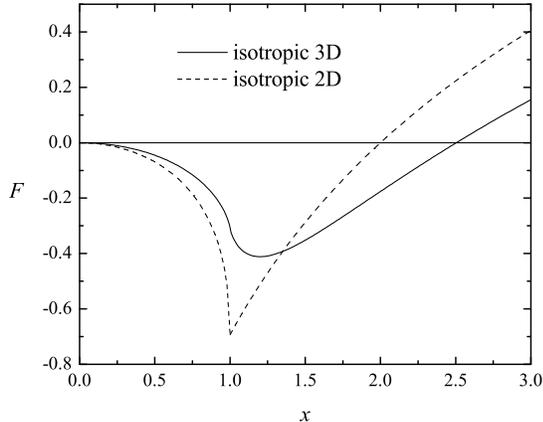}
    \caption{The momentum dependence of ${\cal L}^{-1}(\bq,0)$
    in the isotropic 3D case, Eq. (\ref{FQ 3D}) (solid line), and
    the isotropic 2D case, Eq. (\ref{FQ 2D}) (dashed line); $x=v_F|\bq|/2h$.}
\end{figure}

In the isotropic case the critical field of the LOFF state does
not depend on the direction of $\bq$, and the quantum-critical
fluctuation propagator (\ref{L general}) takes the following form:
\begin{equation}
\label{L 3D isotropic}
    {\cal L}(\bq,\nu_m)=\frac{1}{N_F}\frac{1}{\epsilon+\gamma|\nu_m|+K(|\bq|-q_c)^2},
\end{equation}
where $\gamma\simeq 0.65/h_c$ and $K\simeq 0.28 v_F^2/h_c^2$. The
static limit of Eq. (\ref{L 3D isotropic}) has the same form as
the propagator of classical fluctuations of the order parameter
associated with the crystallization transition in an isotropic
liquid.\cite{Braz75} Note also that the minimum of ${\cal
L}^{-1}(\bq,0)$ remains infinitely degenerate even for an
ellipsoidal Fermi surface, $\xi_{\bk}=\sum_ik_i^2/2m_i-k_F^2/2m$,
where $m_i=m/\mu_i^2$ are the effective masses, $i=x,y,z$. In this
case, a change of variables reduces the fluctuation propagator to
the form (\ref{L 3D isotropic}), in which
$|\bq|\to\sqrt{\mu_x^2q_x^2+\mu_y^2q_y^2+\mu_z^2q_z^2}$.
Therefore, the critical field has the same value as in the
isotropic case, the only difference being that the degeneracy
manifold in the momentum space is now ellipsoidal rather than
spherical.

\subsection{Generic band structure}
\label{sec: generic band}

The infinite degeneracy of the LOFF states is an artifact of the
parabolic band approximation. It will be lifted in the case of a
general band structure or an anisotropic gap symmetry. Since
$A(g^{-1}\bq,h)=A(\bq,h)$, where $g$ is an arbitrary element of
the point group ${\cal G}$, the minima of the inverse fluctuation
propagator in the momentum space form a ``star'', i.e. a set of
$N_q$ isolated points, $\{\bq_c^{(a)}\}$, which is invariant under
all operations from ${\cal G}$. Assuming that inversion is present
in the point group, $N_q$ can be as low as two and as high as the
total number of the group elements in ${\cal G}$. The equilibrium
order parameter just below the critical field can be represented
as a linear combination of the plane waves:
\begin{equation}
\label{LOFF OP}
    \eta(\br)=\sum_{a=1}^{N_q}\eta_ae^{i\bq_c^{(a)}\br}.
\end{equation}
The complex coefficients $\eta_a$, which determine the spatial
structure of the LOFF phase, are found by minimizing the
Ginzburg-Landau free energy. If the minima of $A(\bq,h)$ are
well-separated then the fluctuation modes near different
$\bq_c^{(a)}$ can be treated independently.

Let us consider for concreteness a tetragonal crystal with ${\cal
G}=\mathbf{D}_{4h}$, in which case there can be as many as sixteen
degenerate minima of $A(\bq,h)$. This number is severely reduced
if one assumes that $\bq_c^{(a)}$ are along the highest symmetry
directions. For a 3D band this means that $N_q=2$, with
$\bq_c^{(1,2)}=\pm q_c\hat z$. Near the minima, one can write
$A_{1,2}(\bq,h)=\epsilon+K_\parallel(q_z\mp
q_c)^2+K_\perp(q_x^2+q_y^2)$. On the other hand, if there are 2D
bands in the system, then the lowest possible number of the minima
is four, located for instance at $\bq_c^{(1,3)}=\pm q_c\hat x$,
$\bq_c^{(2,4)}=\pm q_c\hat y$. Then,
$A_1(\bq,h)=\epsilon+K_\parallel(q_x-q_c)^2+K_\perp q_y^2$,
\emph{etc}. Without any loss of generality, we can assume that
$K_\parallel=K_\perp=K$, so that the fluctuation propagator near
the $a$th minimum can be written in the following form:
\begin{equation}
\label{L generic}
     {\cal L}_a(\bq,\nu_m)=\frac{1}{N_F}\frac{1}{\epsilon+\gamma|\nu_m|+
     K\left(\bq-\bq_c^{(a)}\right)^2}.
\end{equation}
This expression, in which $\gamma$ and $K$ should be treated as
phenomenological constants, is applicable in the generic case of a
3D or 2D crystalline superconductor with arbitrary band structure
and pairing symmetry.

We would like to note the formal similarity between our
superconducting fluctuation propagators (\ref{L 3D isotropic}) and
(\ref{L generic}) and the propagators of magnetic fluctuations in
itinerant helical ferromagnets\cite{ST04} and high-$T_c$
superconductors,\cite{MMP90} respectively.

\subsection{Isotropic 2D case}
\label{sec: isotropic 2D}

The fluctuation propagator does not have the simple form (\ref{L
general}) if $\gamma$ is either zero or infinity. According to Eq.
(\ref{tilde FQ}), the former possibility occurs if the surface
$\bv\bQ_c=1$ does not intersect the Fermi surface, or if it does
then $\phi_{\bk}$ accidentally vanishes on the intersection line.
In either case one would have to go to higher orders of the
frequency expansion. We have not been able to find an explicit
example of the band structure for which this happens.

If $\gamma=\infty$ then the Taylor expansion in powers of
$|\nu_m|$ fails, and one to take the low-temperature limit
directly in Eq. (\ref{L}). Assuming as before that the temperature
is the smallest energy scale in the system, one obtains:
\begin{eqnarray}
\label{L with nu}
    &&\frac{1}{N_F}{\cal L}^{-1}(\bq,\nu_m)=\ln\frac{h}{h_0}\nonumber\\
    &&\quad+\re\left\langle\phi_{\bk}^2\ln\left(
    \frac{\bv\bq}{2h}-1-i\frac{|\nu_m|}{2h}\right)
    \right\rangle.
\end{eqnarray}
In order to recover the expressions (\ref{A clean},\ref{tilde A
clean}) from this, one has to replace $|\nu_m|\to|\nu_m|+0^+$.

One can check that the frequency expansion fails in the case of
the isotropic 2D band with $\xi_{\bk}=(k_x^2+k_y^2)/2m-\mu$ and
$\phi_{\bk}=1$. For the order parameter modulated in the $xy$
plane we have
\begin{eqnarray}
\label{FQ 2D}
    &&F(\bQ)=\re\ln\frac{1+\sqrt{1-x^2}}{2},\\
\label{tFQ 2D}
    &&\tilde F(\bQ)=\frac{1}{\sqrt{x^2-1}}\theta(x-1)
\end{eqnarray}
where $x=v_FQ=v_Fq/2h$. The function $F(\bQ)$ has a non-analytical
minimum at $x=1$, with $F(\bQ_c)=-\ln 2$, see Fig. 3. Therefore
the zero-temperature critical field is $h_c=2h_0$, and the LOFF
wave vector is $q_c=4h_0/v_F$. Since $\tilde F$ diverges at the
critical point, one has to use Eq. (\ref{L with nu}), with the
following result:
\begin{equation}
\label{L isotropic 2D}
    \frac{1}{N_F}{\cal L}^{-1}(\bq,\nu_m)=\ln\frac{h}{h_0}+
    {\cal F}\left(\frac{v_F|\bq|}{2h},\frac{|\nu_m|}{2h}\right),
\end{equation}
where
$$
    {\cal F}(x,y)=\re\ln\frac{1+iy+\sqrt{(1+iy)^2-x^2}}{2}.
$$
We see that the retarded fluctuation propagator ${\cal
L}^R(\bq,\nu)$ obtained from Eq. (\ref{L isotropic 2D}) has a
branch cut instead of a simple pole.

The non-analyticity of the inverse fluctuation propagator persists
even if the Fermi surface is a corrugated cylinder. For the
quasi-2D band described by $\xi_{\bk}=(k_x^2+k_y^2)/2m-t\cos
k_zd-\mu$ ($t\ll\mu$), with isotropic pairing, one can show that
the deepest minimum of $A(\bq,h)$ is achieved for
$\bq\parallel\hat z$.\cite{ADF74} Then, $F(\bQ)$ and $\tilde
F(\bQ)$ are given by the same expressions (\ref{FQ 2D}) and
(\ref{tFQ 2D}) as in the 2D case, but with $x=td|q_z|/2h$. The
divergence of $\tilde F$ at $x=1$ again signals the failure of the
expansion of ${\cal L}^{-1}(\bq,\nu_m)$ in powers of frequency.

Below we neglect these complications and assume that the
fluctuation propagator has either the form (\ref{L 3D isotropic})
or (\ref{L generic}). This does not seem to be very restrictive,
especially since a well-defined frequency expansion and analytical
momentum dependence will be restored in realistic layered
superconductors by the Fermi-surface or gap anisotropy, or by
disorder.

\section{Fluctuation corrections}
\label{sec: corrections}

\subsection{Free energy and spin susceptibility}
\label{sec: F}

If the quantum LOFF transition in the isotropic 3D case is first
order,\cite{BR02,MC04} then our theory is not applicable. However,
even in that case it is still instructive to do the calculations
using the fluctuation propagator (\ref{L 3D isotropic}) in order
to highlight the differences with the generic case. The
fluctuation correction to the free energy in the normal state at
$T=0$ is given by
\begin{equation}
\label{F clean}
    \delta F=2\sum_{\bq}\int_0^{\nu_{\max}}\frac{d\nu}{2\pi}\ln{\cal
    L}^{-1}(\bq,\nu).
\end{equation}
The momentum integration is restricted to $|\bq|\leq q_{\max}$,
see Sec. \ref{sec: general}. In addition, the ultraviolet cutoff
$\nu_{\max}\simeq\omega_{\max}$ is introduced to guarantee the
convergence of the frequency integral. This cutoff can be extended
to infinity when calculating the correction to the magnetic
susceptibility:
$$
    \delta\chi=-\frac{\partial^2}{\partial
    H^2}\delta F=-N_F^2\frac{1}{H_c^2}\frac{\partial^2}{\partial\epsilon^2}
    \delta F.
$$

In the isotropic 3D case, using the fluctuation propagator (\ref{L
3D isotropic}), we have at $\epsilon\to 0$:
\begin{eqnarray}
\label{dchi 3D}
    \delta\chi&=&\frac{N_F}{\pi\gamma
    H_c^2}\int\frac{d^3\bq}{(2\pi)^3}\frac{1}{\epsilon+K(|\bq|-q_c)^2}\nonumber\\
    &\simeq&\frac{N_F q_c^2}{2\pi^2\gamma H_c^2\sqrt{K}}
    \frac{1}{\sqrt{\epsilon}}
\end{eqnarray}
(the main contribution to the integral comes from the vicinity of
$q_c\ll q_{\max}$). To estimate the magnitude of the correction,
we compare it to the Pauli spin susceptibility in the normal
state, $\chi_P=2\mu_B^2N_F$:
\begin{equation}
\label{dchi eps 3D}
    \frac{\delta\chi}{\chi_P}\simeq 1.21\left(\frac{\Delta_0}{\epsilon_F}
    \right)^2\left(\frac{H_c}{H-H_c}\right)^{1/2},
\end{equation}
where $\epsilon_F=k_F^2/2m$ is the Fermi energy. Although this
expression is divergent at the quantum critical point, the size of
the fluctuation correction at any nonzero $\epsilon$ is small
because of the factor $\left( \Delta_0/\epsilon_F \right)^2$. The
width of the fluctuation region in this case can be estimated as
$(H-H_c)/H_c \sim  \left( \Delta_0/\epsilon_F \right)^4$. Note
also that the field-dependent fluctuation contribution to the
magnetization,
$$
     \delta M=-\frac{\partial}{\partial H}\delta F
     =-N_F \frac{1}{H_c}\frac{\partial}{\partial\epsilon}\delta F,
$$
is of the type $\delta M\propto\sqrt{\epsilon}$, and is not
singular at $\epsilon\rightarrow 0$.

In the generic 3D case, using Eq. (\ref{L generic}) we obtain:
\begin{equation}
\label{dchi 3D generic}
    \delta\chi=\frac{N_F}{\pi\gamma H_c^2}\sum_a
    \int\frac{d^3\bq}{(2\pi)^3}\frac{1}{\epsilon+K\left(\bq-\bq_c^{(a)}\right)^2}.
\end{equation}
Then, for each minimum $\bq_c^{(a)}$, the integral can be
calculated by shifting the integration variable,
$\bq-\bq_c^{(a)}=\bp$:
\begin{eqnarray}
\label{dchi 3D generic final}
    \delta\chi &=&\frac{N_F}{\pi\gamma H_c^2}\sum_a
    \int\frac{d^3\bp}{(2\pi)^3}\frac{1}{\epsilon+K\bp^2}
    \nonumber\\
    &=&N_q\frac{N_F q_{\max}}{2\pi^3H_c^2\gamma K}
    \left(1-\frac{\pi}{2q_{\max}}\sqrt{\frac{\epsilon}{K}}\right),
\end{eqnarray}
which is not singular at $\epsilon\to 0$, so is the correction to
the magnetization.

In contrast, in the generic 2D case with isolated minima the
reduced dimensionality of the momentum integral leads to a
logarithmic singularity in $\delta\chi$:
\begin{eqnarray}
\label{dchi 2D generic final}
    \delta\chi&=&\frac{N_F}{\pi\gamma H_c^2}\sum_a
    \int\frac{d^2\bp}{(2\pi)^2}\frac{1}{\epsilon+K\bp^2}
    \nonumber\\
    &=&N_q\frac{N_F}{4\pi^2H_c^2\gamma K}
    \ln\frac{Kq_{\max}^2}{\epsilon}.
\end{eqnarray}
Using $\gamma\sim 1/h_c$ and $K\sim v_F^2/h_c^2$, we have the
following estimate for the quantum fluctuation correction to the
susceptibility:
\begin{equation}
\label{dchi eps 2D}
    \frac{\delta\chi}{\chi_P}\sim\frac{\Delta_0}{\epsilon_F}
    \ln\frac{H_c}{H-H_c}.
\end{equation}

Our analysis can be easily extended to the case of the normal
metal-to-uniform superconductor transition (one should remember
though that in a paramagnetically-limited clean superconductor
this transition does not exist, since the LOFF instability always
preempts the uniform instability at $T=0$). Formally setting
$\bq_c=0$ in the fluctuation propagator, we see that the
infinitely degenerate case (\ref{dchi 3D}) is never realized and
the correction to $\chi$ in three dimensions is non-singular, see
Eq. (\ref{dchi 3D generic final}). In the 2D case, one would have
the logarithmically divergent correction (\ref{dchi 2D generic
final}).

\subsection{Quasiparticle decay rate}
\label{sec: Sigma}

The lowest-order contribution to the self-energy of spin-up
fermions due to the superconducting fluctuations in the normal
state is given by
\begin{eqnarray}
\label{Sigma-clean}
    \Sigma_\uparrow(\bk,\omega_n)&=&-T\sum_m\sum_{\bq}
    {\cal L}(\bq,\nu_m)\nonumber\\
    &&\times G_\downarrow\left(-\bk+\bq,-\omega_n+\nu_m\right),
\end{eqnarray}
see Fig. 4, where
$G_\downarrow(\bk,\omega_n)=(i\omega_n-\xi_{\bk}-h)^{-1}$ is the
Green's function of spin-down fermions (the pairing is assumed to
be isotropic). At $T=0$ we obtain for the quasiparticle decay rate
at the spin-up Fermi surface, i.e. for $\bk$ satisfying
$\xi_{\bk}=h$:
\begin{eqnarray}
\label{im Sigma gen}
    \Gamma(\hat\bk,\omega)\equiv-\im\Sigma^R_\uparrow(\hat\bk,\omega)=
    \sum_{\bq}\im{\cal L}^R(\bq,\omega-W_{\bk}),
\end{eqnarray}
where $\hat\bk$ is the direction of the Fermi momentum, ${\cal
L}^R(\bq,\nu)$ is the retarded fluctuation propagator, and the
integration is restricted to the region in the $\bq$-space defined
by the condition $0\leq W_{\bk}\leq\omega$. Below we calculate the
decay rate in the limit $\omega\to 0$ at the quantum critical
point, i.e. at $\epsilon=0$.

\begin{figure}
    \label{fig: Sigma}
    \includegraphics[width=5cm]{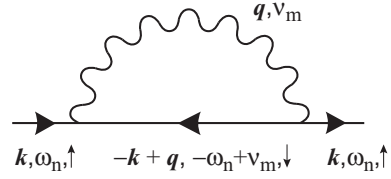}
    \caption{The contribution of superconducting fluctuations to the
    fermion self-energy, Eq. (\ref{Sigma-clean}).
    The solid lines are the single-particle Green's functions.
    The wavy line is the fluctuation propagator ${\cal L}(\bq,\nu_m)$.}
\end{figure}

In the isotropic 3D case the fluctuation propagator is given by
Eq. (\ref{L 3D isotropic}). It is convenient to choose the polar
axis in the $\bq$-space along $\bv=v_F\hat\bk$ (we neglect the
difference between the quasiparticle velocity at the spin-up Fermi
surface and the Fermi velocity $v_F$, which is of the order of
$h_c/\epsilon_F$). It is easy to see that the decay rate in this
case does not depend on $\hat\bk$:
\begin{eqnarray*}
    \Gamma(\omega)&=&\frac{\gamma}{4\pi^2 N_F}\int_{-1}^1ds\\
    &&\times\int_0^\infty q^2dq\frac{\omega-W_{\bk}}{K^2(q-q_c)^4+
    \gamma^2(\omega-W_{\bk})^2}
\end{eqnarray*}
($s=\cos\theta$). Since the main contribution to the integral
comes from from $q\simeq q_c$, one can replace $W_{\bk}\to
v_Fq_cs-2h_c$. Introducing $u=\omega-W_{\bk}$, we have
\begin{eqnarray}
    \Gamma(\omega)&=&
    \frac{\sqrt{2}}{8\pi}\frac{q_c}{ N_F v_F\sqrt{\gamma K}}
    \int_{u_{\min}}^{u_{\max}}du\,u^{-1/2}\nonumber\\
    &=&\frac{\sqrt{2}}{4\pi}\frac{q_c}{N_F v_F\sqrt{\gamma K}}\omega^{1/2},
\end{eqnarray}
where $u_{\min}=\max\{0,\omega-v_Fq_c+2h_c\}=0$,
$u_{\max}=\min\{\omega,\omega+v_Fq_c+2h_c\}=\omega$ (we assume
$\omega\ll h_c$). This expression takes a more transparent form if
compared to the energy scale associated with the superconducting
phase transition:
\begin{equation}
\label{im Sigma 3D}
    \frac{\Gamma(\omega)}{\Delta_0}\simeq
    1.57\left(\frac{\Delta_0}{\epsilon_F}
    \right)^2\left(\frac{\omega}{\Delta_0}\right)^{1/2}.
\end{equation}
We see that at the quantum phase transition into the LOFF state
the Fermi-liquid behavior is destroyed by the superconducting
fluctuations, and the magnitude of the fluctuation contribution to
the decay rate is determined by the factor
$\left(\Delta_0/\epsilon_F\right)^2$.

In the generic case, when the inverse fluctuation propagator has
minima at isolated points $\bq_c^{(a)}$ in the momentum space, see
Eq. (\ref{L generic}), the quasiparticle decay rate can be written
as the sum of the independent contributions from each minimum:
\begin{equation}
\label{Gamma sum}
    \Gamma(\hat\bk,\omega)=\sum_a\Gamma_a(\hat\bk,\omega),
\end{equation}
where
$$
    \Gamma_a=\frac{\gamma}{ N_F}\int\frac{d^D\bq}{(2\pi)^D}
    \frac{\omega-W_{\bk}}{K^2\left(\bq-\bq_c^{(a)}\right)^4+
    \gamma^2(\omega-W_{\bk})^2}.
$$
Here the integration is restricted by the condition $0\leq
W_{\bk}\leq\omega$, and $D=3$ or $2$ is the dimensionality of the
system.

In three dimensions, after changing the variables
$\bq-\bq_c^{(a)}=\bp$, choosing the $z$ axis in the momentum space
along $\bv$, and introducing $u=\omega-w_a-|\bv|p_z$, where
$w_a=\bv\bq_c^{(a)}-2h_c$, one obtains:
\begin{eqnarray*}
    \Gamma_a&=&
    \frac{\gamma}{4\pi^2 N_F|\bv|}\int_0^\infty dp_\perp p_\perp\\
    &&\times\int_0^\omega\frac{du\;u}{K^2[p_\perp^2+(u-\omega+w_a)^2/\bv^2]^2+\gamma^2u^2}.
\end{eqnarray*}
In the limit $\omega\to 0$, one can neglect $u-\omega$ in the
first term in the denominator and calculate the integral over $u$:
\begin{eqnarray}
\label{Gamma_a gen 3D}
    \Gamma_a&=&\frac{1}{8\pi^2N_F\gamma|\bv|}\int_0^\infty dp_\perp p_\perp\nonumber\\
    &&\times\ln\left[1+
    \frac{\gamma^2\omega^2}{K^2(p_\perp^2+w_a^2/\bv^2)^2}\right].\quad
\end{eqnarray}
The result of the integration here essentially depends on whether
$w_a$ is zero or not.

If $\hat\bk$ is such that $w_a\neq 0$, then one can expand the
logarithm in Eq. (\ref{Gamma_a gen 3D}) at $\omega\to 0$ and
calculate the momentum integral. Substituting the result in Eq.
(\ref{Gamma sum}) we obtain
\begin{equation}
\label{Gamma generic FL}
    \Gamma(\hat\bk,\omega)=
    \frac{1}{16\pi^2}\frac{\gamma|\bv|}{N_FK^2\tilde w^2}\omega^2,
\end{equation}
where $\tilde w^{-2}=\sum_aw_a^{-2}$. The fluctuation contribution
to the quasiparticle decay rate has the energy dependence
characteristic of the Fermi liquid, and its magnitude can be
estimated as follows:
\begin{equation}
    \frac{\Gamma(\hat\bk,\omega)}{\Delta_0}\sim
    \left(\frac{\Delta_0}{\epsilon_F}\right)^2
    \left(\frac{\omega}{\Delta_0}\right)^2.
\end{equation}

On the other hand, if, for some $\hat\bk$, one of $w_a$'s is zero,
then $\tilde w^{-2}$ diverges, making the expression (\ref{Gamma
generic FL}) inapplicable. Setting $w_a=0$ in Eq. (\ref{Gamma_a
gen 3D}), one obtains that the decay rate at $\omega\to 0$ is
dominated by the contribution from the $a$th minimum:
\begin{equation}
\label{Gamma generic NFL}
    \Gamma(\hat\bk,\omega)=\frac{1}{16\pi}\frac{1}{N_F|\bv|K}\omega,
\end{equation}
so that
\begin{equation}
    \frac{\Gamma(\hat\bk,\omega)}{\Delta_0}\sim
    \left(\frac{\Delta_0}{\epsilon_F}\right)^2
    \frac{\omega}{\Delta_0}.
\end{equation}
Thus, the following picture emerges: In the generic case, the
energy dependence of the quasiparticle decay rate due to the
interaction with superconducting fluctuations turns out to be
strongly anisotropic on the Fermi surface. While
$\Gamma\propto\omega^2$ is almost everywhere, one has a
non-Fermi-liquid behavior, $\Gamma\propto\omega$, on the lines
defined by the intersection of the surfaces $\bv\bq_c^{(a)}=2h_c$
with the Fermi surface. For the consistency of our calculation we
have to assume that these intersection lines do exist, otherwise
the coefficient $\gamma$ would be zero, see the discussion at the
beginning of Sec. \ref{sec: isotropic 2D}.

Similar conclusions can be obtained in the generic 2D case, in
which the decay rate has the same form (\ref{Gamma sum}), where,
instead of Eq. (\ref{Gamma_a gen 3D}), we now have
\begin{eqnarray}
\label{Gamma_a gen 2D}
    \Gamma_a&=&\frac{1}{8\pi^2 N_F \gamma|\bv|}\int_{-\infty}^\infty dp_\perp\nonumber\\
    &&\times\ln\left[1+
    \frac{\gamma^2\omega^2}{K^2(p_\perp^2+w_a^2/\bv^2)^2}\right].\quad
\end{eqnarray}
If $w_a\neq 0$, then
\begin{equation}
\label{Gamma generic 2D FL}
    \Gamma(\hat\bk,\omega)=\frac{1}{16\pi}
    \frac{\gamma\bv^2}{N_FK^2\tilde w^3}\omega^2,
\end{equation}
where $\tilde w^{-3}=\sum_a|w_a|^{-3}$. Using $\gamma\sim 1/h_c$
and $K\sim v_F^2/h_c^2$, we obtain
\begin{equation}
    \frac{\Gamma(\hat\bk,\omega)}{\Delta_0}\sim
    \frac{\Delta_0}{\epsilon_F}
    \left(\frac{\omega}{\Delta_0}\right)^2.
\end{equation}
At the points on the Fermi surface where $w_a=0$, we have the
expression
\begin{equation}
\label{Gamma generic 2D NFL}
    \Gamma(\hat\bk,\omega)=\frac{\sqrt{2}}{4\pi}
    \frac{1}{ N_F |\bv|\sqrt{\gamma K}}\omega^{1/2},
\end{equation}
whose magnitude can be estimated as
\begin{equation}
    \frac{\Gamma(\hat\bk,\omega)}{\Delta_0}\sim
    \frac{\Delta_0}{\epsilon_F}
    \left(\frac{\omega}{\Delta_0}\right)^{1/2}.
\end{equation}
Similar to the generic 3D case, the fluctuation contribution to
the decay rate strongly depends on the direction of the Fermi
momentum, showing non-Fermi-liquid behavior along some directions.
Although the overall magnitude of the correction is larger than in
the generic 3D case, it is still proportional to the small
parameter $\Delta_0/\epsilon_F$. It is interesting to note that
the same frequency dependence of the decay rate can be obtained in
the model of a nearly antiferromagnetic Fermi liquid in 2D, where
the quasiparticle interaction with spin fluctuations becomes
anomalously strong near some points, the ``hot spots'', on the
Fermi line.\cite{HR95}

Let us now compare our results with the decay rate at the
second-order phase transition into the uniform superconducting
state. Setting $\bq_c=0$ and assuming an isotropic band
dispersion, we have
\begin{equation}
\label{Gamma uniform}
    \Gamma(\omega)=\frac{\gamma}{N_F}\int\frac{d^D\bq}{(2\pi)^D}
    \frac{\omega-W_{\bk}}{K^2\bq^4+\gamma^2(\omega-W_{\bk})^2},
\end{equation}
instead of Eq. (\ref{Gamma sum}). In the 3D case, repeating the
calculation steps leading to Eq. (\ref{Gamma generic FL}), one
obtains:
\begin{equation}
    \Gamma(\omega)=\frac{1}{64\pi^2}
    \frac{\gamma v_F}{N_FK^2\tilde h_c^2}\omega^2.
\end{equation}
Similarly, in the 2D case,
\begin{equation}
    \Gamma(\omega)=\frac{1}{128\pi}
    \frac{\gamma v_F^2}{N_FK^2h_c^3}\omega^2.
\end{equation}
Thus, the Fermi-liquid character of quasiparticle excitations is
not destroyed by the quantum fluctuations at the normal
metal-to-uniform superconductor transition.

\section{Conclusions and outlook}
\label{sec: conclusion}

We have studied the order parameter fluctuations near the quantum
phase transition at $H=H_c$ from the normal state to the LOFF
superconducting state. We derived the general form of the
fluctuation propagator ${\cal L}(\bq,\nu_m)$ at finite $\bq$ and
$\nu_m$. In the systems suggested as good candidates for the
experimental realization of the LOFF state, disorder is small or
absent altogether. In the absence of impurity effects, we analyzed
the momentum and frequency dependence of the fluctuation
propagator in both isotropic 3D and 2D cases, as well as in the
case of generic spectrum.

The fluctuation effects are more pronounced in the isotropic 3D
case compared to the generic situation. This is because in the
isotropic case the LOFF states are infinitely degenerate leading
to the large phase volume of fluctuations. The fluctuation
contribution to the spin susceptibility diverges at $H\to H_c$,
$\delta\chi\propto(H-H_c)^{-1/2}$, and the quasiparticle decay
rate shows a non-Fermi-liquid behavior,
$\Gamma(\omega)\propto\omega^{1/2}$, at the quantum critical
point. The magnitude of the fluctuation corrections is determined
by the parameter $(\Delta_0/\epsilon_F)^2$. Although this ratio is
very small in conventional bulk superconductors, it can vary in a
wide range in the atomic Fermi gases. It should be noted that our
results rely on the assumption that the LOFF transition at $T=0$
is of second order. As mentioned in the Introduction, this might
not be the case in clean isotropic systems in the weak-coupling
limit. However, as the ratio $\Delta_0/\epsilon_F$ grows so do the
strong-coupling corrections to the mean-field free energy, which
could stabilize the second-order LOFF transition. More theoretical
work is needed to check if this possibility can indeed be
realized.

In the generic case, which is expected to be applicable to
crystalline paramagnetically-limited superconductors, or to the
atomic Fermi gases in optical lattices, the equilibrium wave
vectors of the LOFF state form a set of isolated points in the
momentum space. The phase volume of fluctuations is reduced,
resulting in a non-singular spin susceptibility in the 3D case,
and a weak, logarithmic, divergence of $\delta\chi$ in the 2D
case. Interestingly, the fluctuations in the generic case are
still strong enough to cause the breakdown of the Fermi liquid at
the quantum critical point, which manifests itself in a highly
anisotropic energy dependence of the quasiparticle decay rate on
the Fermi surface: $\Gamma(\omega)\propto\omega$ on some lines in
3D, and $\Gamma(\omega)\propto\omega^{1/2}$ at some points in 2D.
We expect that the Fermi-liquid behavior,
$\Gamma(\omega)\propto\omega^2$, will be restored throughout the
Fermi surface away from the quantum critical point or in the
presence of disorder, leaving the details for a future
publication.

There is a number of open questions concerning the assumptions we
made and the effects we neglected in the present study, the order
of the quantum phase transition into the LOFF state being
particularly important. Even if the mean-field transition is of
second order, this might no longer be the case if one takes into
account the fluctuation renormalization of the free energy.
Additional complications arise in realistic crystalline
superconductors, in which the orbital effects and disorder should
be included. Finally, it would be interesting to extend our
calculation of the fluctuation corrections to nonzero temperatures
in the critical region around the LOFF transition.

\section*{Acknowledgements}

This work was supported by the Natural Sciences and Engineering
Research Council of Canada.

\end{document}